\documentclass[12pt]{article}
\usepackage{amsmath} \usepackage{amsfonts} \usepackage{amssymb}\usepackage{cite}

\newtheorem{prop}{Proposition}

\newtheorem{lem}{Lemma}

\newtheorem{rem}{Remark}

\def\Cb{{\mathbb C}}
\def\qed{\relax\ifmmode\hskip2em \Box\else\unskip\nobreak\hskip1em $\Box$\fi}
\newfont{\uno}{cmr12 scaled 1100}
\newcommand{\one}{\uno{1}\hspace{-3.3pt}\mathrm{l}}

\begin{document}
\centerline{\Large{\bf Multipartite states under local unitary
transformations}} \vspace{4ex}
\begin{center}
Sergio Albeverio$^{a}$ \footnote{SFB 611; IZKS; BiBoS;
CERFIM(Locarno); Acc. Arch. USI (Mendrisio)

~~~e-mail: albeverio@uni-bonn.de}, ~Laura
Cattaneo$^{a}$ \footnote{e-mail: cattaneo@wiener.iam.uni-bonn.de},
 ~Shao-Ming Fei$^{a,b,c}$
\footnote{e-mail: fei@uni-bonn.de}, ~Xiao-Hong
Wang$^{b}$ \footnote{e-mail: wangxh@mail.cnu.edu.cn}

\vspace{2ex}
\begin{minipage}{4.8in}

{\small $~^{a}$ Institut f\"ur Angewandte Mathematik,
Universit\"at Bonn, D-53115}

{\small $~^{b}$ Department of Mathematics, Capital Normal
University, Beijing 100037}

{\small $~^{c}$ Max Planck Institute for
Mathematics in the Sciences, 04103 Leipzig}

\end{minipage}
\end{center}

\vskip 1 true cm
\parindent=18pt
\parskip=6pt

\vspace{0.5cm}
\begin{abstract}
The equivalence problem under local unitary transformation for $n$--partite pure states is reduced to the one for
$(n-1)$--partite mixed states. In particular, a tripartite system
$\mathcal{H}_A\otimes\mathcal{H}_B\otimes\mathcal{H}_C$, where $\mathcal{H}_j$ is a finite dimensional complex
Hilbert space for $j=A,B,C$, is considered and a set of invariants under local transformations is introduced,
which is complete for the set of states whose partial trace with respect to $\mathcal{H}_A$ belongs to the class
of generic mixed states.\\[0.3cm]
\end{abstract}

\subsection*{Introduction}
The importance of a measure to quantify entanglement became
evident in the years by the number of applications exploiting
nonlocality properties which have been developed: we mention,
among others, quantum computation (see, e.g.,
\cite{Intro1,Intro2}), quantum teleportation (see, e.g.,
\cite{Intro3,Intro4,Intro5,Intro6,Intro7,Intro8,Intro9,Intro10}),
superdense coding (see, e.g., \cite{Intro11}), quantum
cryptography (see, e.g., \cite{Intro12,Intro13,Intro14}).\\

Many proposals have been made for a measure of entanglement in the
bipartite case, see e.g.,
\cite{bi1,bi2,bi3,bi4,bi5,AlFePaYa03,wang,bi6}. Less results are
known instead for the tripartite and in general for the
$n$--partite case \cite{AlFePaYa03,tri1,tri2,AlCaFeWa04}, although
such systems are important for example in quantum multipartite
teleportation or
telecloning processings.\\

One of the properties employed in the bipartite case is the
Schmidt-decomposition \cite{Sc1907}. However this decomposition is
a peculiarity of bipartite systems and does not exist for
$n$--partite ones, a sign of the complexity of the many-partite
problem. Generalizations of the Schmidt-decomposition have been
proposed \cite{ds1,ds2,ds3,ds4}, but the results are not
sufficient to provide good measures of entanglement in the
$n$--partite case. In the following, we first reduce the
$n$--partite problem to a ($n$-$1$)--partite one. To illustrate
this, we consider the case of a tripartite system. Then we define
invariants under local unitary transformations which form a complete set at least for tripartite
states for which a solution of the bipartite problem for
entanglement measures is known.

\subsection*{Tripartite states as bipartite ones}
Let $\mathcal{H}_A$, $\mathcal{H}_B$, and $\mathcal{H}_C$ be
complex Hilbert spaces of finite dimension $N_A$, $N_B$, and
$N_C$, respectively, and let $\{|j\rangle_k\}_{j=1}^{N_k}$,
$k=A,B,C$, be an orthonormal basis of $\mathcal{H}_k$. A pure
state $|\psi\rangle$ in
$\mathcal{H}_A\otimes\mathcal{H}_B\otimes\mathcal{H}_C$ can then
be written as
\[|\psi\rangle = \sum_{j=1}^{N_A}\sum_{k=1}^{N_B}\sum_{l=1}^{N_C}a_{jkl}|j\rangle_A\otimes|k\rangle_B\otimes|l\rangle_C\,,
\quad\quad\sum_{j=1}^{N_A}\sum_{k=1}^{N_B}\sum_{l=1}^{N_C}a_{jkl}a_{jkl}^*=1\,.
\]
We denote by $\mathrm{U(}\mathcal{H}\mathrm{)}$ the group of all
unitary operators on the space $\mathcal{H}$. \\

First of all, we can consider tripartite states as special cases
of bipartite ones, by decomposing the system into two subsystems,
for example $A$--$BC$. The following lemma holds.
\begin{lem}
Let $|\psi\rangle$, $|\psi'\rangle$ be two pure states in
$\mathcal{H}_A\otimes\mathcal{H}_B\otimes\mathcal{H}_C$ and define
$\rho=\operatorname{Tr}_A\left(|\psi\rangle\langle\psi|\right)$,
$\rho'=\operatorname{Tr}_A\left(|\psi'\rangle\langle\psi'|\right)$,
where $\operatorname{Tr}_A$ denotes the partial trace with respect
to $\mathcal{H}_A$.
\begin{itemize}
\item [a)] The function $I_{\alpha}^A(|\psi\rangle) =
\operatorname{Tr}\rho^{\alpha}$ is invariant under local unitary
transformations, for any $\alpha\in\mathbb{N}$; \item [b)] If
$I_{\alpha}^A(|\psi'\rangle)=I_{\alpha}^A(|\psi\rangle)$ for
$\alpha =1,\dots, \operatorname{min}\{N_A, N_B\cdot  N_C\}$, there
exist $U_A\in\mathrm{U(}\mathcal{H}_A\mathrm{)}$,
$U_{BC}\in\mathrm{U(}\mathcal{H}_B\otimes\mathcal{H}_C\mathrm{)}$
such that $|\psi'\rangle=U_A\otimes U_{BC}|\psi\rangle$. In
particular, $\rho'=U_{BC}\rho U_{BC}^{\dagger}$.
\end{itemize}
\label{l1}
\end{lem}
\noindent {\bf Proof.} As already shown in \cite{AlCaFeWa04}, {\it a)} is easily proved as
Tr$_A(|\psi\rangle\langle\psi|)=A_A^TA_A^*$, where $A_A$ is the matrix obtained considering $|\psi\rangle$ as a
bipartite state in the $A$--$BC$ system, with the row (resp. column) indices from the subsystem $A$ (resp. $BC$).
The indices $^T$ resp. $^*$ denote transpose resp. complex conjugation. As an example,
\[A_A=\left(\begin{array}{cccc}  a_{111} & a_{112} & a_{121} & a_{122}\\ a_{211} & a_{212} & a_{221}
& a_{222}\end{array}\right)\] is the matrix  $A_A$ for the case
$N_A=N_B=N_C=2$. Indeed, if $|\psi'\rangle=U_A\otimes U_B\otimes
U_C|\psi\rangle$, with $U_i\in\mathrm{U(}\mathcal{H}_i\mathrm{)}$,
$i=A,B,C$, then $A_A'$ and $A_A$ are related by
\[A_A'=U_A A_A(U_B\otimes U_C)^T\]
and
\begin{eqnarray*}
I_{\alpha}^A(|\psi'\rangle) &=& \operatorname{Tr}(A_A'^T
A_A'^*)^{\alpha}=\operatorname{Tr}((U_A
A_A(U_B\otimes U_C)^T)^T(U_A A_A(U_B\otimes U_C)^T)^*)^{\alpha}\\
&=& \operatorname{Tr}(U_B\otimes
U_C(A_A^TA_A^*)^{\alpha}(U_B\otimes
U_C)^{\dagger})=\operatorname{Tr}(A_A^TA_A^*)^{\alpha}\\
&=&I_{\alpha}^A(|\psi\rangle)
\end{eqnarray*}
for any power $\alpha\in\mathbb{N}$. The decomposition
$|\psi'\rangle=U_A\otimes U_{BC}|\psi\rangle$ follows directly
considering $|\psi\rangle$ as a bipartite state of the system
$A$--$BC$ and applying the results of \cite{wang}. \qed\\

\begin{rem}\em
The statement can be generalized to $n$--partite systems: the equivalence problem for $n$--partite pure
states is reduced in this way to the equivalence problem for $(n-1)$-partite mixed states.
\end{rem}

\subsection*{Reduction to bipartite mixed states}
Lemma \ref{l1} allows us to reduce the tripartite problem on
$\mathcal{H}_A\otimes\mathcal{H}_B\otimes\mathcal{H}_C$  to a
bipartite problem on $\mathcal{H}_B\otimes\mathcal{H}_C$.
\begin{lem}
Let $|\psi'\rangle=U_A\otimes U_{BC}|\psi\rangle$, with
$U_A\in\mathrm{U(}\mathcal{H}_A\mathrm{)}$,
$U_{BC}\in\mathrm{U(}\mathcal{H}_B\otimes\mathcal{H}_C\mathrm{)}$
and define
$\rho=\operatorname{Tr}_A\left(|\psi\rangle\langle\psi|\right)$,
$\rho'=\operatorname{Tr}_A\left(|\psi'\rangle\langle\psi'|\right)$.
If
\[\rho'=U_B\otimes U_C\rho U_{B}^{\dagger}\otimes U_{C}^{\dagger}\,,\]
where $U_B\in\mathrm{U(}\mathcal{H}_B\mathrm{)}$ and
$U_C\in\mathrm{U(}\mathcal{H}_C\mathrm{)}$, then there exist
matrices $V_A\in\mathrm{U(}\mathcal{H}_A\mathrm{)}$,
$V_B\in\mathrm{U(}\mathcal{H}_B\mathrm{)}$,
$V_C\in\mathrm{U(}\mathcal{H}_C\mathrm{)}$ such that
\[|\psi'\rangle=V_A\otimes V_B\otimes V_C|\psi\rangle\,,\]
i.e., $|\psi\rangle$ and $|\psi'\rangle$ are equivalent under
local unitary transformations. \label{l2}
\end{lem}
\noindent {\bf Proof.} On one hand we have
\begin{eqnarray*}
U_{BC}\operatorname{Tr}_A\left(|\psi\rangle\langle\psi|\right)^{\alpha}U_{BC}^{\dagger}&=&
\operatorname{Tr}_A\left(\one\otimes
U_{BC}|\psi\rangle\langle\psi| (\one\otimes
U_{BC})^{\dagger}\right)^{\alpha}\\ &=&
\operatorname{Tr}_A\left(U_A \otimes
U_{BC}|\psi\rangle\langle\psi| (U_A \otimes
U_{BC})^{\dagger}\right)^{\alpha}\,,
\end{eqnarray*}
on the other hand
\[U_B\otimes U_C\operatorname{Tr}_A
\left(|\psi\rangle\langle\psi|\right)^{\alpha}U_{B}^{\dagger}\otimes
U_{C}^{\dagger}= \operatorname{Tr}_A\left(U_A\otimes U_B\otimes
U_C|\psi\rangle\langle\psi| (U_A\otimes U_B\otimes
U_C)^{\dagger}\right)^{\alpha}\,.\] Since this holds for any power
$\alpha\in\mathbb{N}$, there exist a local unitary transformation
$W_A$ on $\mathcal{H}_A$ such that
\begin{eqnarray*}
\lefteqn{U_A \otimes U_{BC}|\psi\rangle\langle\psi| (U_A \otimes U_{BC})^{\dagger}}\\
&&= (W_A\otimes\one\otimes\one) U_A\otimes U_B\otimes
U_C|\psi\rangle\langle\psi| (U_A\otimes U_B\otimes
U_C)^{\dagger}(W_A\otimes\one\otimes\one)^{\dagger}\\
&&= W_AU_A\otimes U_B\otimes U_C|\psi\rangle\langle\psi|
(W_AU_A\otimes U_B\otimes U_C)^{\dagger}\,.
\end{eqnarray*}
Hence
\[|\psi'\rangle=U_A \otimes U_{BC}|\psi\rangle=
\widetilde{U}_A\otimes U_B\otimes U_C |\psi\rangle\,,\] where
$\widetilde{U}_A$ is equal $W_AU_A$ up to a phase factor. \qed\\

Lemma \ref{l1} and Lemma \ref{l2} together give rise to the
following proposition.
\begin{prop}
For pure states $|\psi\rangle$ and $|\psi'\rangle$,
$\rho=\operatorname{Tr}_A\left(|\psi\rangle\langle\psi|\right)$
and
$\rho'=\operatorname{Tr}_A\left(|\psi'\rangle\langle\psi'|\right)$,
we have that
$I_{\alpha}^A(|\psi'\rangle)=I_{\alpha}^A(|\psi\rangle)$ for
$\alpha =1,\dots, \operatorname{min}\{N_A, N_B\cdot  N_C\}$ and
$\rho'=U_B\otimes U_C\rho U_B^{\dagger}\otimes U_C^{\dagger}$ for
some $U_B\in\mathrm{U(}\mathcal{H}_B\mathrm{)}$,
$U_C\in\mathrm{U(}\mathcal{H}_C\mathrm{)}$, if and only if
$|\psi\rangle$ and $|\psi'\rangle$ are equivalent under local
unitary transformations. \label{p1}
\end{prop}

\begin{rem}\em A result corresponding to Lemma \ref{l1},
Lemma \ref{l2}, and Proposition \ref{p1} holds when tripartite is replaced by $n$--partite, for any $n\geqslant
3$, by splitting the system $A_1A_2\dots A_n$ into, e.g., $A_1$--$A_2\dots A_n$.
\end{rem}

\subsection*{New invariants}
The next step is to find further invariants under local unitary
transformations which give the same value for two states if and
ony if $\rho'$ can be written as $U_B\otimes U_C\rho
U_B^{\dagger}\otimes U_C^{\dagger}$ for some unitary
transformations $U_B\in\mathrm{U(}\mathcal{H}_B\mathrm{)}$,
$U_C\in\mathrm{U(}\mathcal{H}_C\mathrm{)}$, the main obstacle
being the fact that in general $\rho$ is a bipartite mixed state
and there is no general characterization of entanglement for that case. \\

The generalization of $I_{\alpha}^A(|\psi\rangle)$ to bipartite
mixed states is
$\operatorname{Tr}(\operatorname{Tr}_j(\rho))^{\alpha}$, where
$j=B,C$. For a pure state
$|\psi\rangle\in\mathcal{H}_A\otimes\mathcal{H}_B\otimes\mathcal{H}_C$
this means to consider the functions
\[\operatorname{Tr}(\operatorname{Tr}_j
\left(\operatorname{Tr}_A|\psi\rangle\langle\psi|\right))^{\alpha}\,.\]
Therefore we introduce the following set of new invariants
\begin{equation}
I_{\alpha,\beta}^{j,k}(|\psi\rangle) =
\operatorname{Tr}(\operatorname{Tr}_k\left(\operatorname{Tr}_j|
\psi\rangle\langle\psi|\right)^{\alpha})^{\beta}\,, \label{Inv}
\end{equation}
where $j,k\in\{A,B,C\}$, $j\neq k$, and
$\alpha,\beta\in\mathbb{N}$ .
\begin{lem}
The functions $I_{\alpha,\beta}^{j,k}(|\psi\rangle)$ defined in
(\ref{Inv}) are invariant under local unitary transformations
$U_A\otimes U_B\otimes U_C$.
\end{lem}
\noindent {\bf Proof.} As a model we consider $I_{\alpha,\beta}^{A,B}(|\psi\rangle)$. The other cases can be
treated in an analogous manner. We have
\begin{equation}
\operatorname{Tr}_A(|\psi\rangle\langle\psi|)
=\sum_{j=1}^{N_A}\sum_{k,p=1}^{N_B}\sum_{l,q=1}^{N_C}
a_{jkl}a_{jpq}^*|kl\rangle\langle pq|\,, \label{Inv4}
\end{equation}
where $|kl\rangle$ stands for $|k\rangle_B\otimes|l\rangle_C$.
Multiplying (\ref{Inv4}) $\alpha$ times ($\alpha\in\mathbb{N}$)
and calculating the partial trace on $\mathcal{H}_B$ of the matrix
obtained we get
\begin{eqnarray*}
\lefteqn{\operatorname{Tr}_B\left(\operatorname{Tr}_A|\psi\rangle\langle\psi|\right)^{\alpha}}
\\
&=& \sum^{N_A}_{\substack{j_{1}=1, \\ j_{2}=1, \\\dots, \\
j_{\alpha}=1}}\; \sum^{N_B}_{\substack{p_{1}=1, \\ p_{2}=1, \\
\dots,\\ p_{\alpha}=1}}\; \sum^{N_C}_{\substack{q_{1}=1, \\
q_{2}=1, \\ \dots,\\ q_{\alpha}=1,\\m_{1}=1}}
a_{j_{1}p_{1}q_{1}}^* a_{j_{2}p_{1}q_{1}}a_{j_{2}p_{2}q_{2}}^*
a_{j_{3}p_{2}q_{2}}\dots a_{j_{\alpha}p_{\alpha-1}q_{\alpha-1}}
a_{j_{\alpha}p_{\alpha}q_{\alpha}}^* a_{j_{1}p_{\alpha}m_{1}}
|m_{1}\rangle\langle q_{\alpha}|
\end{eqnarray*}
and hence
\begin{eqnarray*}
\lefteqn{\operatorname{Tr}(\operatorname{Tr}_B\left(\operatorname{Tr}_A
|\psi\rangle\langle\psi|\right)^{\alpha})^{\beta}}\\
&=&
\prod^{\beta}_{k=1}\Big(\sum^{N_A}_{\substack{j_{k_1}=1, \\
j_{k_2}=1, \\\dots, \\ j_{k_{\alpha}}=1}}\;
\sum^{N_B}_{\substack{p_{k_1}=1, \\ p_{k_2}=1, \\ \dots,\\
p_{k_{\alpha}}=1}}\; \sum^{N_C}_{\substack{q_{k_1}=1, \\
q_{k_2}=1, \\ \dots,\\ q_{k_{\alpha}}=1}}
a_{j_{k_1}p_{k_1}q_{k_1}}^* a_{j_{k_2}p_{k_1}q_{k_1}}\dots
a_{j_{k_{\alpha}}p_{k_{\alpha-1}}q_{k_{\alpha-1}}}
a_{j_{k_{\alpha}}p_{k_{\alpha}}q_{k_{\alpha}}}^*a_{j_{k_1}p_{k_{\alpha}}q_{{k-1}_{\alpha}}}\Big)\,,
\end{eqnarray*}
where $q_{0_{\alpha}}\equiv q_{\beta_{\alpha}}$. Instead of the
one employed in the proof of Lemma \ref{l1}, an alternative way to
consider the factors $a_{jkl}$ is by writing them in matrices
$(A^{(j)})_{kl}$: the index $j$ sets the considered matrix and
$k$, $l$ describe the row and column of $A^{(j)}$, respectively.
That is, we write $|\psi\rangle =
\sum_{j=1}^{N_A}\sum_{k=1}^{N_B}\sum_{l=1}^{N_C}
A^{(j)}_{kl}|jkl\rangle$. Using this notation, one obtains
\begin{eqnarray*}
\lefteqn{\hspace{-3cm}\operatorname{Tr}(\operatorname{Tr}_B\left(\operatorname{Tr}_A
|\psi\rangle\langle\psi|\right)^{\alpha})^{\beta}}\\
= \sum^{N_A}_{\substack{j_{1_1},\dots,j_{1_{\alpha}}=1 \\
j_{2_1},\dots,j_{2_{\alpha}}=1\\ \dots \\
\\j_{\beta_1},\dots,j_{\beta_{\alpha}}=1}} &&
\Big(\prod^{\beta}_{k=1}\operatorname{Tr}({A^{(j_{k_1})}}^{\dagger}A^{(j_{k_2})})
\operatorname{Tr}({A^{(j_{k_2})}}^{\dagger}A^{(j_{k_3})}) \dots
\operatorname{Tr}({A^{(j_{k_{\alpha-1}})}}^{\dagger}A^{(j_{k_{\alpha}})})\Big)\\
&&\cdot\operatorname{Tr}({A^{(j_{\beta_{\alpha}})}}^{\dagger}A^{(j_{\beta_{1}})}
{A^{(j_{{\beta-1}_{\alpha}})}}^{\dagger}A^{(j_{{\beta-1}_{1}})}\dots
{A^{(j_{1_{\alpha}})}}^{\dagger}A^{(j_{1_{1}})})\,.
\end{eqnarray*}
For a local unitary transformations $U\otimes V\otimes W$ we have
\begin{eqnarray*}
|\psi'\rangle \!\! &:=&\!\!\! U\otimes V\otimes W|\psi\rangle
=\sum_{j=1}^{N_A}\sum_{k=1}^{N_B}\sum_{l=1}^{N_C}
{A'}^{(j)}_{kl}|jkl\rangle\\
U\otimes V\otimes W|\psi\rangle \!\! &=&\!\!\!\!\!
\sum_{j,m=1}^{N_A}\sum_{k, p=1}^{N_B}\sum_{l,q=1}^{N_C}
{A}^{(j)}_{kl}U_{mj}V_{pk}W_{ql}|mpq\rangle=\!\sum_{j,m=1}^{N_A}\sum_{k=1}^{N_B}\sum_{l=1}^{N_C}
U_{jm}(VA^{(m)}W^T)_{kl}|jkl\rangle,
\end{eqnarray*}
i.e., ${A'}^{(j)}_{kl}=\sum_{m=1}^{N_A} U_{jm}(VA^{(m)}W^T)_{kl}$
and
\[\operatorname{Tr}({{A'}^{(p_{q_r})}}^{\dagger}{A'}^{(p_{q_{r+1}})}) = \sum^{N_A}_{m_1,m_2=1}U^{\dagger}_{m_1p_{q_r}}U_{p_{q_{r+1}}m_2}
\operatorname{Tr}({A^{(m_1)}}^{\dagger}A^{(m_2)}).\] Therefore
\begin{eqnarray*}
&&\hspace{-0.9cm}\Big(\prod^{\beta}_{k=1}\operatorname{Tr}({{A'}^{(j_{k_1})}}^{\dagger}{A'}^{(j_{k_2})})
\operatorname{Tr}({{A'}^{(j_{k_2})}}^{\dagger}{A'}^{(j_{k_3})})
\dots
\operatorname{Tr}({{A'}^{(j_{k_{\alpha-1}})}}^{\dagger}{A'}^{(j_{k_{\alpha}})})\Big)\cdot\\
&&\hspace{-0.6cm}\cdot
\operatorname{Tr}({{A'}^{(j_{\beta_{\alpha}})}}^{\dagger}{A'}^{(j_{\beta_{1}})}
{{A'}^{(j_{{\beta-1}_{\alpha}})}}^{\dagger}{A'}^{(j_{{\beta-1}_{1}})}\dots
{{A'}^{(j_{1_{\alpha}})}}^{\dagger}{A'}^{(j_{1_{1}})})\\
&&\hspace{-0.5cm}
=\sum^{N_A}_{\substack{m_{1_1},\dots,m_{1_{\alpha}}=1 \\
m_{2_1},\dots,m_{2_{\alpha}}=1\\ \dots \\
m_{\beta_1},\dots,m_{\beta_{\alpha}}=1 \\
p_{1_{\alpha}},\dots,p_{\beta_{\alpha}}=1}}
\sum^{N_A}_{\substack{n_{1_1},\dots,n_{1_{\alpha}}=1\\
n_{2_1},\dots,n_{2_{\alpha}}=1\\ \dots \\
n_{\beta_1},\dots,n_{\beta_{\alpha}}=1\\q_{1_1},\dots,q_{\beta_1}=1}}
\Big(\prod^{\beta}_{k=1}
U^{\dagger}_{m_{k_1}j_{k_1}}U_{j_{k_2}n_{k_2}}U^{\dagger}_{m_{k_2}j_{k_2}}U_{j_{k_3}n_{k_3}}\dots
U_{j_{k_{\alpha}}n_{k_{\alpha}}}\\
&&\hspace{4.7cm}\cdot\operatorname{Tr}({A^{(m_{k_1})}}^{\dagger}A^{(n_{k_2})})
\operatorname{Tr}({A^{(m_{k_2})}}^{\dagger}A^{(n_{k_3})}) \dots
\operatorname{Tr}({A^{(m_{k_{\alpha-1}})}}^{\dagger}A^{(n_{k_{\alpha}})})\Big)\\
&&\hspace{4cm}\cdot
U^{\dagger}_{p_{\beta_{\alpha}}j_{\beta_{\alpha}}}U_{j_{\beta_1}q_{\beta_1}}
U^{\dagger}_{p_{{\beta-1}_{\alpha}}j_{{\beta-1}_{\alpha}}}U_{j_{{\beta-1}_1}q_{{\beta-1}_1}}\dots
U_{j_{1_1}q_{1_1}}\\[0.2cm]
&&\hspace{4cm}\cdot\operatorname{Tr}({A^{(p_{\beta_{\alpha}})}}^{\dagger}A^{(q_{\beta_{1}})}
{A^{(p_{{\beta-1}_{\alpha}})}}^{\dagger}A^{(q_{{\beta-1}_{1}})}\dots
{A^{(p_{1_{\alpha}})}}^{\dagger}A^{(q_{1_{1}})})\,.
\end{eqnarray*}
The result follows, since $U$ is unitary and hence $\sum_k
U^{\dagger}_{jk}U_{kl}=\delta_{jl}$. \qed\\

\begin{rem}\em The invariants
$I_{\alpha,\beta}^{j,k}(|\psi\rangle)$ can easily be generalized to $n$--partite systems: the functions
\[I_{\alpha_1,\alpha_2,\dots,\alpha_n}^{j_1,j_2,\dots,j_n}
(|\psi\rangle)=\operatorname{Tr} \left(\operatorname{Tr}_{j_1}
\left(\operatorname{Tr}_{j_2}\left(\dots\left(
\operatorname{Tr}_{j_n}| \psi\rangle\langle\psi|\right)^{\alpha_n}
\dots\right)^{\alpha_3}\right)^{\alpha_2}  \right)^{\alpha_1}\,,
\quad\alpha_i\in\mathbb{N}\,,\quad i=1,\dots,n\,,\]
are invariant under local unitary transformations $U_1\otimes U_2\otimes\dots\otimes U_n$.
\end{rem}

Unfortunately, the invariants (\ref{Inv}) seem to be sufficient
only in the case in which the $\lambda_j$ of the decomposition
$\rho=\sum^{n}_{j=1}\lambda_j|\varphi_j\rangle\langle\varphi_j|$,
where $n\leqslant N_B\cdot N_C$ and
$\varphi_j\in\mathcal{H}_B\otimes\mathcal{H}_C$ for all $j$, are
not degenerated, i.e, $\lambda_j\neq\lambda_k$ for $j\neq k$.
Indeed, the following lemma holds.
\begin{lem}
Let $|\psi\rangle$ and $|\psi'\rangle$ be two tripartite pure
states such that
$I_{\alpha,\beta}^{j,k}(|\psi\rangle)=I_{\alpha,\beta}^{j,k}(|\psi'\rangle)$
for $j,k\in\{A,B,C\}$ and $j\neq k$, $\alpha=1,\dots, N_q\cdot
N_r$, and $\beta=1,\dots, N_r$, where $q, r\in\{A,B,C\}$ and $r$ is different from $j$, $k$ and $q$. Then,
\begin{itemize}
\item [a)] there exist $U_p\in\mathrm{U(}\mathcal{H}_p\mathrm{)}$
and
$U_{q,r}\in\mathrm{U(}\mathcal{H}_q\otimes\mathcal{H}_r\mathrm{)}$,
with $p$, $q$, $r$ different from each other, such that
$|\psi'\rangle = U_p\otimes U_{q,r}|\psi\rangle$; \item [b)] for
any $|\varphi_m\rangle$ of the decomposition
$\operatorname{Tr}_p\left(|\psi\rangle\langle\psi|\right)
=\sum^{n}_{m=1}\lambda_m^{(p)}|\varphi_m^{(p)}\rangle\langle
\varphi_m^{(p)}|$ for which $\lambda_m^{(p)}$ is not degenerate we
have
\[U_{q,r}|\varphi_m^{(p)}\rangle=v_q^m\otimes
u_r|\varphi_m^{(p)}\rangle=u_q\otimes
v_r^m|\varphi_m^{(p)}\rangle\,,\] where $v_q^m,
u_q\in\mathrm{U(}\mathcal{H}_q\mathrm{)}$ and $u_r,
v_r^m\in\mathrm{U(}\mathcal{H}_r\mathrm{)}$.
\end{itemize}
\label{l3}
\end{lem}
\noindent {\bf Proof.} Part a) was already proved in Lemma \ref{l1}, since
\[I_{\alpha}^p(|\psi\rangle)=I_{\alpha,1}^{p,k}(|\psi\rangle)=I_{\alpha,1}^{p,k}(|\psi'\rangle)= I_{\alpha}^{p}(|\psi'\rangle)\,.\]
Further we know that
$\operatorname{Tr}_p\left(|\psi'\rangle\langle\psi'|\right)
 = U_{q,r}\operatorname{Tr}_p\left(|\psi\rangle\langle\psi|\right)U_{q,r}^{\dagger}$.
Since
$I_{\alpha,\beta}^{i,k}(|\psi'\rangle)=I_{\alpha,\beta}^{i,k}(|\psi\rangle)$
for $\beta=1,\dots, N_r$, with $r$ different from $i$ and $k$, there exists a
$u_r\in\mathrm{U(}\mathcal{H}_r\mathrm{)}$ such that
\[\operatorname{Tr}_k\left(\operatorname{Tr}_i|\psi'\rangle\langle\psi'|\right)^{\alpha} = u_r\operatorname{Tr}_k\left(\operatorname{Tr}_i|\psi\rangle\langle\psi|\right)^{\alpha}u_r^{\dagger}\,.\]
Therefore, since this result holds for all $\alpha=1,\dots,
N_q\cdot N_r$, where $i$, $q$, $r$ are different from each other,
and the $\lambda_m^{(p)}$ are not degenerated, for any $m$ there
exists a $u_q^{(m)}\in\mathrm{U(}\mathcal{H}_q\mathrm{)}$ such
that
\begin{eqnarray*}
U_{q,r}|\varphi_m^{(p)}\rangle\langle\varphi_m^{(p)}|U_{q,r}^{\dagger}
&=& (u_q^{(m)}\otimes\one)(\one\otimes u_r)
|\varphi_m^{(p)}\rangle\langle\varphi_m^{(p)}|(\one\otimes
u_r)^{\dagger}(u_q^{(m)}\otimes\one)^{\dagger}\\
&=& (u_q^{(m)}\otimes
u_r)|\varphi_m^{(p)}\rangle\langle\varphi_m^{(p)}|(u_q^{(m)}\otimes
u_r)^{\dagger}\,.
\end{eqnarray*}
The statement follows, as $|\langle\varphi_m^{(p)}|(u_q^{(m)}\otimes
u_r)^{\dagger}U_{q,r}|\varphi_m^{(p)}\rangle|=1$. \qed

\begin{rem}\em
\begin{enumerate}
\item Lemma \ref{l3} b) is only sufficient if $\rho$ is a pure
state. \item For $n$-partite pure states the condition
$I_{\alpha,\beta}^{j,k}
(|\psi\rangle)=I_{\alpha,\beta}^{j,k}(|\psi'\rangle)$ for $j,
k\in\{A_1,A_2,\dots, A_n\}$ implies $|\psi'\rangle = U_{p_1}
\otimes U_{p_2,p_3,\dots,p_n}|\psi\rangle$ for some
$U_{p_1}\in\mathrm{U(}\mathcal{H}_{p_1}\mathrm{)}$,
$U_{p_2,p_3,\dots,p_n}\in\mathrm{U(}\mathcal{H}_{p_2}\otimes\dots\otimes\mathcal{H}_{p_n}\mathrm{)}$
and
\[U_{p_2,\dots,p_n}|\varphi_j^{(p_1)}\rangle = v_{p_2}^j\otimes u_{p_3,\dots,
p_n}|\varphi_j^{(p_1)}\rangle\] for any
$|\varphi_j^{(p_1)}\rangle$ of the decomposition
$\operatorname{Tr}_{p_1}\left(|\psi\rangle\langle\psi|\right)
=\sum^{n_1}_{j=1}\lambda_j^{(p_1)}|\varphi_j^{(p_1)}\rangle\langle\varphi_j^{(p_1)}|$
such that $\lambda_j^{(p_1)}$ is not degenerated. Further
\[u_{p_3,\dots,p_n}^j|\varphi_{j,k}^{(p_2)}\rangle = v_{p_3}^{j,k}\otimes u_{p_4,\dots,
p_n}^j|\varphi_{j,k}^{(p_2)}\rangle\] for
$\operatorname{Tr}_{p_2}\left(|\varphi_j^{(p_1)}\rangle\langle\varphi_j^{(p_1)}|\right)
=\sum^{n_2}_{k=1}\lambda_{j,k}^{(p_2)}|\varphi_{j,k}^{(p_2)}\rangle\langle\varphi_{j,k}^{(p_2)}|$,
if  $\lambda_{j,k}^{(p_2)}$ and $\lambda_j^{(p_1)}$ are not
degenerated, and so on. Note that only the invariants
$I_{\alpha,\beta}^{j,k}$ were considered, and not
$I_{\alpha_1,\alpha_2,\dots,\alpha_n}^{j_1,j_2,\dots,j_n}$.
\end{enumerate}
\end{rem}

\subsection*{A special case for tripartite states}
Complete sets of invariants for the case of bipartite mixed states
are known only for some special cases. For example, in
\cite{wang} a complete set was presented for the case in
which the state $\rho=\sum_{m=1}^{n}\lambda_m|\varphi_m\rangle\langle\varphi_m|$ is a generic mixed state. To define this set,
we need further invariants:
\[\Theta(\rho)_{jk}=\operatorname{Tr}\left(\operatorname{Tr}_B
(|\varphi_j\rangle\langle\varphi_j|)^* \operatorname{Tr}_B
(|\varphi_k\rangle\langle\varphi_k|)^*\right)\,,\quad\Omega(\rho)_{jk}
=\operatorname{Tr}\left(\operatorname{Tr}_C(|\varphi_j\rangle\langle\varphi_j|)
\operatorname{Tr}_C(|\varphi_k\rangle\langle\varphi_k|)\right)\,.\] Assume without loss of generality that
$N_B\leqslant N_C$ and complete $\Theta(\rho)$ and $\Omega(\rho)$ to $(N_B^2\times N_B^2)$-matrices by defining
$\Theta(\rho)_{jk}=\Omega(\rho)_{jk}=0$ for $n<j,k\leqslant N_B^2$. A bipartite mixed state is called generic if
the $(N_B^2\times N_B^2)$-matrices $\Theta(\rho)$ and $\Omega(\rho)$ are non-degenerate.\\

If $\rho$ is a generic mixed state and $U\rho U^{\dagger}$, with $U$ unitary, gives the same values as $\rho$ for
the invariants $J_{\alpha}^j(\rho)=\operatorname{Tr}(\operatorname{Tr}_j(\rho^{\alpha}))$, where $j\in\{B,C\}$,
$\Theta(\rho)$, $\Omega(\rho)$, and
\begin{eqnarray*}
Y(\rho)_{jkl}&=&\operatorname{Tr}\left(\operatorname{Tr}_B(|\varphi_j\rangle\langle\varphi_j|)^*
\operatorname{Tr}_B(|\varphi_k\rangle\langle\varphi_k|)^*
(\operatorname{Tr}_B(|\varphi_l\rangle\langle\varphi_l|)^*\right)\,,\\
X(\rho)_{jkl}&=&\operatorname{Tr}\left(\operatorname{Tr}_C(|\varphi_j\rangle\langle\varphi_j|)
\operatorname{Tr}_C(|\varphi_k\rangle\langle\varphi_k|)
(\operatorname{Tr}_C(|\varphi_l\rangle\langle\varphi_l|)\right)\,,
\end{eqnarray*}
where $j,k,l=1,\dots, n$, then $\rho$ and $U\rho U^{\dagger}$ are equivalent under local unitary transformations
\cite{wang}. That is, if $\operatorname{Tr}_A(|\psi\rangle\langle\psi|)$ is a generic mixed state and the above
invariants give the same results for $\operatorname{Tr}_A(|\psi\rangle\langle\psi|)$ and
$\operatorname{Tr}_A(|\psi'\rangle\langle\psi'|)$, as well as
$I_{\alpha}^A(|\psi\rangle)=I_{\alpha}^A(|\psi'\rangle)$ for $\alpha=1,\dots,\operatorname{min}\{N_A,N_B^2\}$,
$|\psi\rangle$ and $|\psi'\rangle$ are equivalent under local unitary transformations. The number of invariants
one needs to calculate can be diminished if one considers (\ref{Inv}) and takes into account Lemma \ref{l3}.
\begin{prop}
Let $|\psi\rangle$ and $|\psi'\rangle$ be two pure
states of $\mathcal{H}_A\otimes\mathcal{H}_B\otimes\mathcal{H}_C$ and assume that $\rho = \operatorname{Tr}_A(|\psi\rangle\langle\psi|)$
is a generic mixed state. $|\psi\rangle$ is equivalent to
$|\psi'\rangle$ under local unitary transformations if and only if
\begin{equation}
I_{\alpha,\beta}^{A,s}(|\psi\rangle)=I_{\alpha,\beta}^{A,s}(|\psi'\rangle)
\label{Inv3}
\end{equation}
for $s\in\{B,C\}$, $\alpha=1,\dots,\operatorname{min}\{N_B^2,N_C^2\}$, $\beta=1,\dots,N_r$,
where $r\in\{B,C\}$ but is different from $s$, and for $\rho' =
\operatorname{Tr}_A(|\psi'\rangle\langle\psi'|)$
\begin{equation}
\Theta(\rho)_{jk}
=\Theta(\rho')_{jk}\,,\quad\Omega(\rho)_{jk}=\Omega(\rho')_{jk}\,,\quad
Y(\rho)_{jkl}=Y(\rho')_{jkl}\,,\quad X(\rho)_{jkl}=X(\rho')_{jkl}
\label{Inv2}
\end{equation}
for the $j,k$ such that $\lambda_j=\lambda_k$.
\label{prop2}
\end{prop}
\noindent {\bf Proof.} As remarked above, the invariants (\ref{Inv2}) are sufficient to establish whether two
states for which the partial trace on $\mathcal{H}_A$ is a generic mixed state are equivalent or not. It remains
to prove that (\ref{Inv2}) is fulfilled when $\lambda_j$, $\lambda_k$, and $\lambda_l$ are non-degenerate, if
(\ref{Inv3}) holds. This follows from Lemma \ref{l3} Indeed, for example
\begin{eqnarray*}
\operatorname{Tr}_C(|\varphi'_j\rangle\langle\varphi'_j|) &=&
\operatorname{Tr}_C(U_{BC}|\varphi_j\rangle\langle\varphi_j|U_{BC}^{\dagger})
= \operatorname{Tr}_C(u_B\otimes
v_C^j|\varphi_j\rangle\langle\varphi_j|(u_B\otimes
v_C^j)^{\dagger})\\
&=& u_B\operatorname{Tr}_C(\one\otimes
v_C^j|\varphi_j\rangle\langle\varphi_j|(\one\otimes
v_C^j)^{\dagger})u_B^{\dagger}=u_B\operatorname{Tr}_C(|\varphi_j\rangle\langle\varphi_j|)u_B^{\dagger}\,,
\end{eqnarray*}
hence
\begin{eqnarray*}
\Omega(\rho')_{jk} &=&
\operatorname{Tr}\left(\operatorname{Tr}_C(|\varphi'_j\rangle\langle\varphi'_j|)
\operatorname{Tr}_C(|\varphi'_k\rangle\langle\varphi'_k|)\right)
=\operatorname{Tr}(u_B\operatorname{Tr}_C(|\varphi_j\rangle\langle\varphi_j|)u_B^{\dagger}u_B
\operatorname{Tr}_C(|\varphi_k\rangle\langle\varphi_k|)u_B^{\dagger})\\
&=&
\operatorname{Tr}\left(\operatorname{Tr}_C(|\varphi_j\rangle\langle\varphi_j|)
\operatorname{Tr}_C(|\varphi_k\rangle\langle\varphi_k|)\right)
=\Omega(\rho)_{jk}\,.
\end{eqnarray*}
The same holds for $\Theta(\rho)$, $Y(\rho)$, and $X(\rho)$.\qed\\

\begin{rem}\em
We know that the rank of $\rho$ is smaller than $\operatorname{min}\{N_A,N_B\cdot N_C\}$ (see, e.g., \cite{DuViCi1100}).
On the other hand, the assumption that $\rho$ is a generic mixed state implies that
$\rho$ has full rank, i.e., $N_B\cdot N_C$. Therefore, in order to fulfill the
conditions of Proposition \ref{prop2}, we need $N_A\geqslant N_B\cdot
N_C$.
\end{rem}

In this last section we have seen that a criterion for equivalence of a class of
bipartite mixed states gives rise to a criterion of
equivalence for a class of pure tripartite states. In \cite{ds5},
the complete invariants for another two classes of bipartite mixed
states are given. For bipartite mixed states on $\Cb^m \times
\Cb^n$,
$$
\rho=\sum_{l=0}^N \mu_l |\xi_l\rangle\langle\xi_l|,
$$
where the rank of $\rho$ is $N+1$ ($N \geq 1$), $\mu_l$ are
eigenvalues with corresponding eigenvectors $|\xi_l\rangle
=\sum_{ij}\xi^{(l)}_{ij}|ij\rangle$. Let $A_l:=(\xi^{(l)}_{ij}),$
$\rho_l:=A_lA_l^*$, and $\theta_l:=A_l^*A_l$, for $l=0,1,...,N$.
If each eigenvalue of $\rho_0$ and $\theta_0$ has multiplicity one
(i.e., is ``multiplicity free''), then $\rho$ belongs to the class
of density matrices to which a complete set of invariants can be
explicitly given. For rank two mixed states on $\Cb^m \times
\Cb^n$ such that each of the matrices $\rho_0$, $\rho_1$,
$\theta_0$, and $\theta_1$ has at most two different eigenvalues,
an operational criterion can be also found. From these criteria
for bipartite mixed states, by using Lemma \ref{l3} we can
similarly obtain criteria for some classes of pure tripartite
states.

\subsection*{Conclusion}
We have reduced the equivalence problem for $n$--partite pure
states to the one for ($n$-$1$)--partite mixed states and in the
special case $n=3$ we have constructed a set of invariants under local unitary
transformations which is complete for the states with partial
trace on $\mathcal{H}_A$ which is a generic mixed state.  \\

\noindent {\bf Acknowledgments.} The second named author
gratefully acknowledges the financial support by the Stefano
Franscini Fund. The fourth author gratefully acknowledges the
support provided by the China-Germany Cooperation Project 446 CHV
113/231, ``Quantum information and related mathematical problems".

\end{document}